\documentclass[%
 aip,
 amsmath,amssymb,
reprint,%
]{revtex4-1}

\usepackage{graphicx}
\usepackage{dcolumn}
\usepackage{bm}

\usepackage{upgreek}
\usepackage{hyperref}
\usepackage{xcolor}
\hypersetup{
    colorlinks = true,
    linkcolor  = {black},
    citecolor  = {[rgb]{0.3,0.49,0.62}},
    urlcolor = {[rgb]{0.3,0.49,0.62}}
}

\usepackage[utf8]{inputenc}
\usepackage[T1]{fontenc}
\usepackage{mathptmx}

\begin{document}

\title{Field-free switching of magnetic tunnel junctions driven by spin-orbit torques at sub-ns timescales}

\author{Viola Krizakova}
    \email{viola.krizakova@mat.ethz.ch}
     \affiliation{Department of Materials, ETH Zurich, 8093 Zurich, Switzerland.}
\author{Kevin Garello}%
    \affiliation{imec, Kapeldreef 75, 3001 Leuven, Belgium}%
\author{Eva Grimaldi}
     \affiliation{Department of Materials, ETH Zurich, 8093 Zurich, Switzerland.}
\author{Gouri Sankar Kar}%
    \affiliation{imec, Kapeldreef 75, 3001 Leuven, Belgium}%
\author{Pietro Gambardella}
   \email{pietro.gambardella@mat.ethz.ch}
    \affiliation{Department of Materials, ETH Zurich, 8093 Zurich, Switzerland.}%

\date{\today}

\begin{abstract}
We report time-resolved measurements of magnetization switching by spin-orbit torques in the absence of an external magnetic field in perpendicularly magnetized magnetic tunnel junctions (MTJ). Field-free switching is enabled by the dipolar field of an in-plane magnetized layer integrated above the MTJ stack, the orientation of which determines the switching polarity. Real-time single-shot measurements provide direct evidence of magnetization reversal and switching distributions. Close to the critical switching voltage we observe stochastic reversal events due to a finite incubation delay preceding the magnetization reversal. Upon increasing the pulse amplitude to twice the critical voltage the reversal becomes quasi-deterministic, leading to reliable bipolar switching at sub-ns timescales in zero external field.
We further investigate the switching probability as a function of dc bias of the MTJ and external magnetic field, providing insight on the parameters that determine the critical switching voltage.  
\end{abstract}

\maketitle

\begin{figure}
\includegraphics{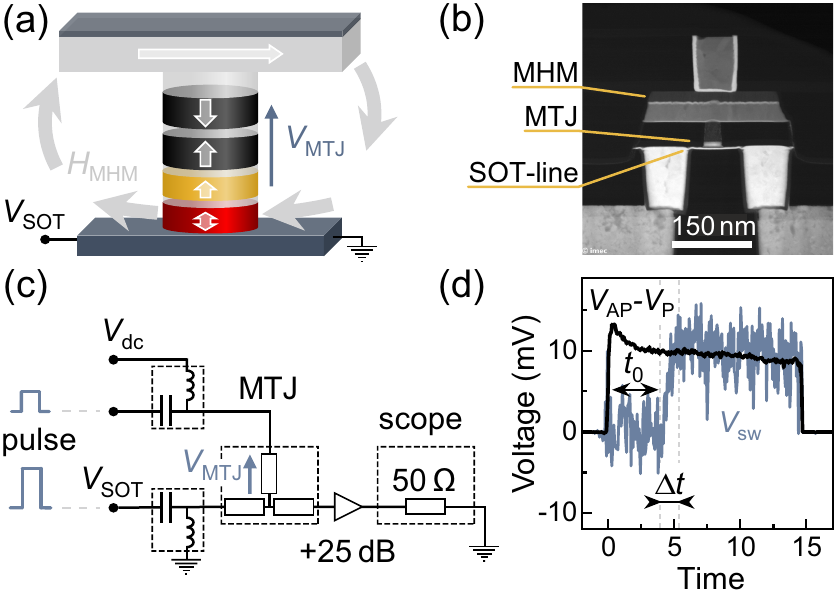}
\caption{\label{fig1:SampleSetup} (a) Schematic and (b) TEM cross-sectional view of a field-free switching MTJ device. The field $H_\text{MHM}$ produced by the MHM is represented by gray arrows. (c) Simplified schematic of the measurement circuit for after-pulse and real-time detection of SOT and/or STT switching. (d) Voltage difference between P and AP states measured by the scope and averaged over 300 switching events (black line). Single-shot switching voltage trace (blue line) prior to normalization (see text), and definition of the characteristic times $t_0$ and $\Delta t$. The positive direction of the bias voltage $V_\text{MTJ}$ is indicated in (a) and (c).}
\end{figure}

Current-induced spin-orbit torques (SOT) allow for manipulating the magnetization of diverse classes of magnetic materials and devices \cite{Manchon2019,Miron2011,Liu2012,Cubukcu2014,Fukami2016b,Lee2016,Luo2020}. Recent studies on ferromagnetic nanodots and magnetic tunnel junctions (MTJ) have shown that SOT-induced switching can overcome spin transfer torque (STT) switching in terms of speed, reliability, and endurance \cite{Garello2014,Zhang2015,Aradhya2016,Baumgartner2017,Yoon2017,Decker2017,Cubukcu2018,Kato2018a,Grimaldi2020}. Fast and reliable deterministic switching of MTJs is especially important for the development of non-volatile magnetic random access memories \cite{Kent2015,Apalkov2016,Prenat2016}.
However, when switching a perpendicular magnetization by SOT, a static in-plane magnetic field is required to break the torque symmetry, which otherwise does not discriminate between up and down magnetic states \cite{Miron2011}. As such field is detrimental for memory applications, various approaches have been proposed to achieve field-free SOT switching, including exchange coupling to an antiferromagnet \cite{Fukami2016,Oh2016,VanDenBrink2016}, RKKY \cite{Lau2016} and Dzyaloshinskii-Moriya \cite{Luo2019} coupling to a reference ferromagnet, tilted magnetic anisotropy \cite{You2015}, geometrical asymmetry \cite{Yu2014a,Safeer2016,Chen2018a}, and two-pulse schemes \cite{VanDenBrink2014,Wang2018,DeOrio2019}. An alternative approach, compatible with back-end-of-line integration of MTJs on CMOS wafers, is based on embedding a ferromagnet in the hard mask that is used to pattern the SOT current line \cite{Garello2019}. In such devices, the rectangular shape of the magnetic hard mask (MHM) provides strong shape anisotropy along the current direction, thus generating an in-plane field on the free layer without imposing additional overheads on the processing and power requirements of the MTJs [Fig.~\ref{fig1:SampleSetup}(a)].

In this letter, we report on real-time measurements of field-free magnetization switching in three-terminal MTJs including a MHM. Whereas the switching dynamics has been extensively studied in two-terminal MTJs operated by STT \cite{Devolder2008,Tomita2008,Cui2010,Hahn2016,Devolder2016,Devolder2016a}, there are only few studies addressing the transient dynamics and real-time reversal speed of individual SOT-induced switching events in three-terminal MTJs \cite{Grimaldi2020,Inokuchi2019}. In particular, the SOT-induced dynamics, reversal speed, and critical voltage in MTJs with perpendicular magnetization have not been investigated in the absence of an external magnetic field.

\begin{figure*}
\includegraphics{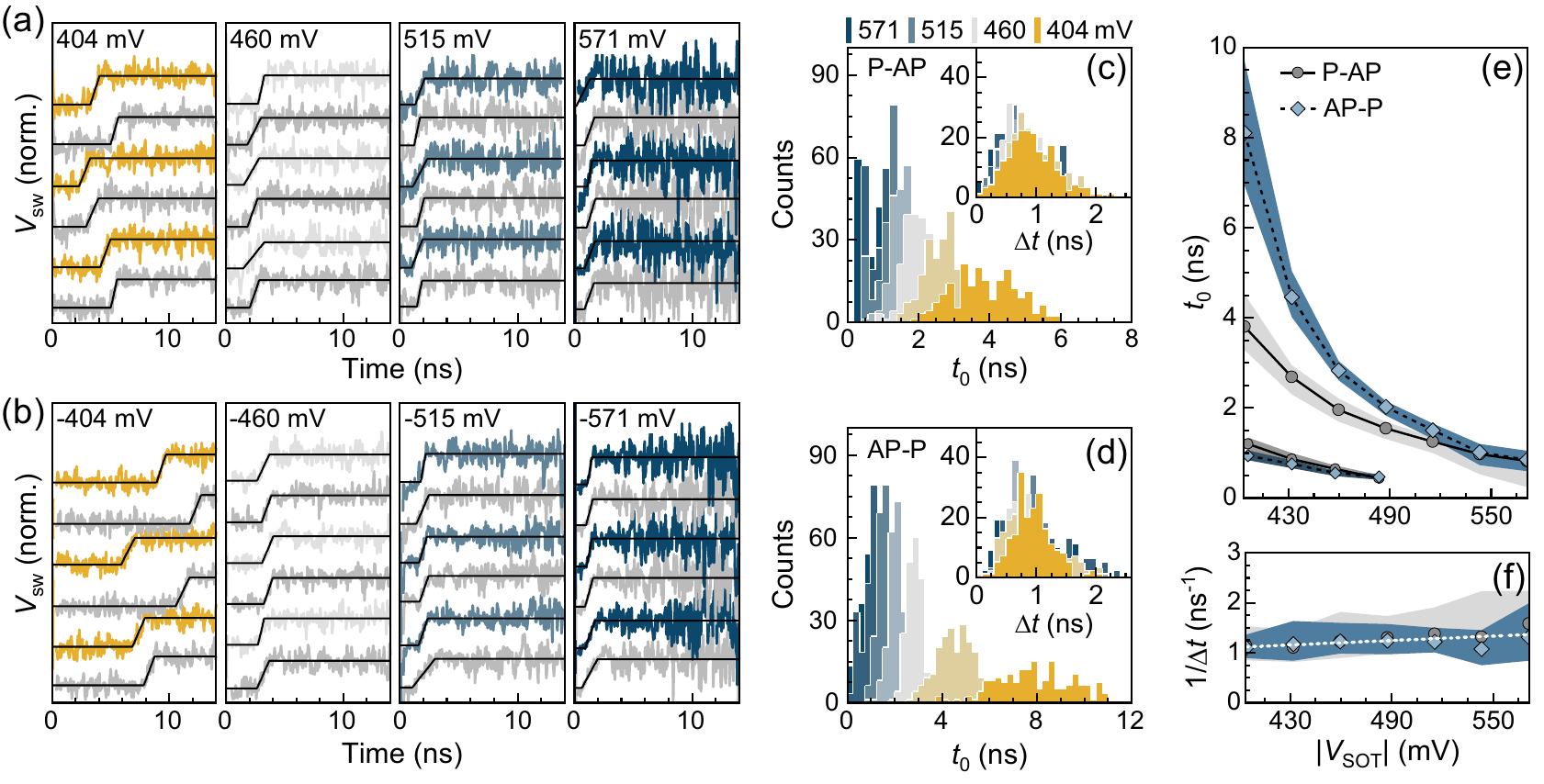}
\caption{\label{fig2:TRsingles} Representative single-shot time traces for different pulse amplitudes of $V_\text{SOT}$ acquired at zero external field for (a) P-AP and (b) AP-P switching at $V_\text{MTJ}$\,=\,-0.5$V_\text{SOT}$. The characteristic times $t_0$ and $\Delta t$ are extracted by fitting the data to a linear ramp (black lines). (c) Statistical distribution of $t_0$ and $\Delta t$ for P-AP and (d) AP-P switching. (e) Dependence of $t_0$ and (f) $1/\Delta t$ on $V_\text{SOT}$ extracted from the distributions. Symbols denote the median values, shaded areas represent the lower and upper quartiles of the distributions. The bottom curves in (e) show $t_0$ measured with $V_\text{MTJ} = -1.65V_\text{SOT}$. The dashed line in (f) is a linear fit to $1/\Delta t$.}
\end{figure*}
Our devices are top-pinned MTJs with 108\% tunneling magnetoresistance (TMR) patterned into a circular pillar with a diameter of 80\,nm. The pillars are grown on top of a 190\,nm wide $\beta$-W SOT-current line with resistivity of $160\,\upmu\Omega$\,cm and resistance of $370\,\Omega$. The MTJ is formed by free and reference layers made of CoFeB with a thicknesses of 0.9 and 1\,nm, respectively, separated by an MgO tunnel barrier with resistance-area product of 20\,$\Omega\,\upmu$m$^2$, as shown in Fig.~\ref{fig1:SampleSetup}(a). The reference layer (yellow) is pinned to a synthetic antiferromagnet (black, SAF). The reference layer and the SAF generate an out-of-plane dipolar field $\approx 13$\,mT that favors the anti-parallel (AP) over the parallel (P) state of the MTJ. The hard mask used to pattern the SOT line incorporates a 50\,nm thick Co magnet\cite{Garello2019}, which provides an in-plane field ($\mu_0 H_\text{MHM}\approx 40$\,mT) parallel to the SOT line [Figs.~\ref{fig1:SampleSetup}(a) and (b)]. Due to its high aspect ratio of $130\times 410$\,nm$^2$, the magnetization direction of the Co hard mask remains constant after saturation. Moreover, the magnetization of the hard mask is not influenced by pulsing the current either through the SOT line or the MTJ pillar.

Electrical measurements are performed using a setup that combines real-time and after-pulse readout of the MTJ resistance, as depicted in Fig.~\ref{fig1:SampleSetup}(c) and reported in more detail in Ref.\,\onlinecite{Grimaldi2020}. After-pulse switching measurements consist of an initialization pulse that sets the free layer magnetization in the desired state, which is verified by a dc measurement of the MTJ resistance, and a switching pulse, also followed by a dc resistance measurement. In the time-resolved measurements, a driving voltage supplied by a pulse generator with 0.15\,ns rise time is split in two pulses with a well-defined amplitude that are simultaneously fed to the input electrodes of the three-terminal MTJ device. The pulse applied to the bottom electrode ($V_\text{SOT}$) drives the SOT reversal. Together, the pulse applied to the top electrode and $V_\text{SOT}$ determine the potential difference across the MTJ pillar ($V_\text{MTJ}$). By adjusting the ratio between these pulse amplitudes, we control the instantaneous value of $V_\text{MTJ}$, which allows for studying phenomena emerging from the bias during SOT-driven reversal, such as voltage control of magnetic anisotropy (VCMA) as well as STT switching \cite{Grimaldi2020}. In this work, we restrict ourselves to switching at zero, low, and strong bias ($V_\text{MTJ}$\,=\,0, -0.5 and $\text{-1.65}$\,$V_\text{SOT}$). The pulse transmitted through the device is amplified and acquired on a 20\,GHz sampling oscilloscope, which allows for monitoring the MTJ resistance in real-time in a time window defined by the width of the driving pulse. To facilitate the analysis of different switching events, a reference trace is subtracted to each voltage trace recorded during a switching pulse. The reference trace is obtained by maintaining the MTJ in its initial state, either P or AP, by application of an external magnetic field opposing $H_\text{MHM}$. The resultant trace is then divided by the voltage difference between the P and AP states [Fig.~\ref{fig1:SampleSetup}(d)] in order to obtain the normalized switching signal $V_\text{sw}$.

Time-resolved studies of current-induced switching can be performed either by pump-probe measurements \cite{Baumgartner2017,Decker2017,Yoon2017}, which yield the average magnetization dynamics, or single-shot measurements \cite{Devolder2008,Tomita2008,Cui2010,Hahn2016,Devolder2016,Devolder2016a,Inokuchi2019,Grimaldi2020}. Although average time-resolved measurements provide information on the reproducible dynamic behavior of the magnetization and afford a higher signal-to-noise ratio compared to single-shot measurements, stochastic dynamical processes can only be revealed in studies carried out on individual switching events. In this study we perform both types of measurements to highlight different aspects of the reversal dynamics. In order to provide a measurable tunneling magnetoresistance reading in single-shot measurements, we apply a small voltage bias $V_\text{MTJ}$ on the MTJ to allow for current flow through the pillar. Depending on the sign of $V_\text{MTJ}$ relative to $V_\text{SOT}$, this bias can either assist or hinder the SOT switching. Two effects are induced by $V_\text{MTJ}$, namely the STT and VCMA. As the sign of the STT depends on the orientation of the reference layer and the VCMA does not, these two effects can be disentangled from each other \cite{Grimaldi2020}. Here, we set the sign of $V_\text{MTJ}$ such that STT always opposes SOT switching for the chosen orientation of the reference layer and $H_\text{MHM}$. We thus focus primarily on SOT-induced switching, unlike previous work in which STT was used to promote switching \cite{Grimaldi2020,Wang2018}. Moreover, in our configuration, the VCMA balances the effect of the SAF dipolar field.

Figures\,\ref{fig2:TRsingles}(a) and \ref{fig2:TRsingles}(b) show representative time traces of individual P-AP and AP-P switching events obtained in zero external field for different pulse amplitudes of $V_\text{SOT}$  and $V_\text{MTJ}\text{ = -0.5\,} V_\text{SOT}$, which corresponds to a current density $j_\text{MTJ} < 30\%$ of the critical STT switching current. The switching traces reveal that the reversal of the free layer starts after a finite incubation time ($t_0$) followed by a single jump of the resistance during a relatively short transition time ($\Delta t$), after which the magnetization remains quiescent in the final state until the pulse ends. The reversal dynamics is thus qualitatively similar to that observed in the presence of an external magnetic field \cite{Grimaldi2020}. Accordingly, we attribute $t_0$ to the time required to nucleate a reversed domain and $\Delta t$ to the time to propagate a domain wall across the free layer \cite{Grimaldi2020,Baumgartner2017}. Noise in the time traces noticeably increases upon increasing $V_\text{SOT}$ and with the time elapsed from the pulse onset, which we associate with the rise of the device temperature during the pulse. Each reversal trace is fit by a linear ramp and the characteristic times $t_0$ and $\Delta t$, defined in Fig.\,\ref{fig1:SampleSetup}(d), are extracted from the line breakpoints. The distributions of $t_0$ and $\Delta t$ representing statistics over 200 single-shot measurements are plotted in Figs.\,\ref{fig2:TRsingles}(c) and \ref{fig2:TRsingles}(d). At low pulse amplitude, typical values of $t_0$ significantly exceed $\Delta t$. However, both the center and the width of the $t_0$ distribution can be reduced by more than one order of magnitude by increasing $V_\text{SOT}$.

To investigate the characteristic times in more detail, we extract the median as well as the lower and upper quartiles of each distribution and plot them as a function of $|V_\text{SOT}|$. Figure\,\ref{fig2:TRsingles}(e) shows that the median $t_0$ decreases to below 1\,ns for both AP-P and P-AP switching when increasing $V_\text{SOT}$ from 400 to 560\,mV. At the lowest pulse amplitudes, the median $t_0$ of the AP-P switching configuration is twice as long compared to P-AP switching. This asymmetry, which is attributed to the dipolar field of the SAF, gradually reduces upon increasing $V_\text{SOT}$. Additionally, our measurements show that such an asymmetry can be strengthened or eliminated by tuning $V_\text{MTJ}$. To demonstrate the potential of SOT-driven switching at increased $V_\text{MTJ}$, we report in the same plot $t_0$ obtained at $V_\text{MTJ}\text{ = -1.65\,}V_\text{SOT}$ (bottom curve). In this case, the difference between both configurations is minimized by VCMA, which favors AP-P at the expense of P-AP switching, whereas the median value and its dispersion are reduced by the bias-induced temperature rise in the device.  Note that STT has the same hindering effect on both switching configurations, as it favors the orientation of the free layer opposite to the final state defined by $V_\text{SOT}$. Moreover, despite the presence of a strong opposing STT when $V_\text{MTJ}\text{ = -1.65\,}V_\text{SOT}$, which corresponds to $j_\text{MTJ} < 96\%$ of the critical STT switching current in the absence of SOT, we do not observe writing errors within our data set.

In contrast to $t_0$, $\Delta t$ has a much weaker dependence on $V_\text{SOT}$. Figure\,\ref{fig2:TRsingles}(f) shows that $1/\Delta t$ increases linearly with $V_\text{SOT}$, with a moderate slope of 1.5\,ns$^{-1}$\,V$^{-1}$. Linear scaling with current is indeed expected for the speed of SOT-driven domain walls in the flow regime \cite{Miron2011NatMat,Thiaville2012,Emori2013,Martinez2013}. Supposing that the reversal initiates with the nucleation of a domain wall at one edge of the free layer, as shown in previous work \cite{Mikuszeit2015,Baumgartner2017}, our data imply an average domain-wall propagation speed of 100\,m\,s$^{-1}$ induced by an SOT current density of $1.7\times 10^{12}$\,A\,m$^{-2}$ at $V_\text{SOT}$\text{ = 0.48\,V}. 

\begin{figure}
\includegraphics{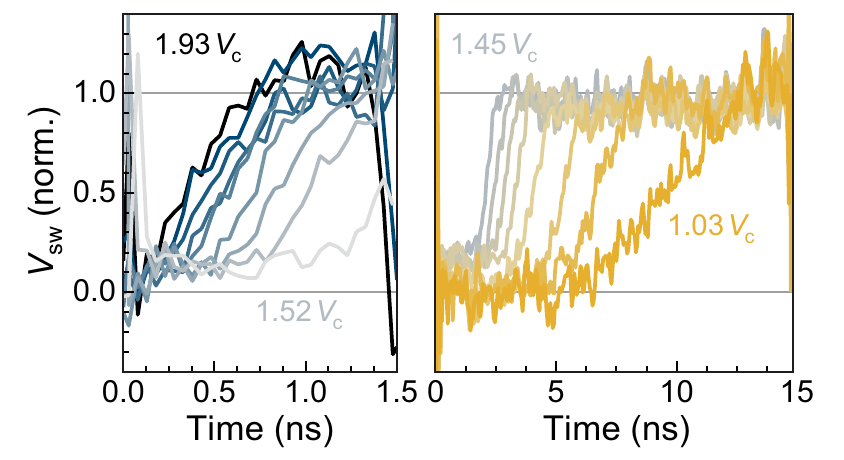}
\caption{\label{fig3:TRave} Averaged time traces of AP-P switching for 1.5-ns (left) and 15-ns-long (right) pulses at different $V_\text{SOT}$ pulse amplitudes close to $V_\text{c}$ acquired at $V_\text{MTJ} = 0.1 V_\text{SOT}$ and zero external field. P-AP switching traces (not shown) yield similar results.}
\end{figure}
To demonstrate the reliability of switching for repeated events, we performed time-resolved measurements averaged over 1000 switching trials, as shown in Fig.\,\ref{fig3:TRave}. Averaging the acquired waveforms allows us to decrease the bias down to $V_\text{MTJ}\text{ = 0.1\,}V_\text{SOT}$, which corresponds to a current density $j_\text{MTJ}$ that is $<15$\% of the STT switching threshold. The time traces compare AP-P switching for different values of $V_\text{SOT}$ given in multiples of the critical switching voltage ($V_\text{c}$), corresponding to 50\% switching probability. Each trace comprises an initial delay and a smooth transition part without noticeable intermediate states. Shortening of the delay and transition part of the averaged time traces indicates that the switching process changes from stochastic to almost deterministic upon increasing $V_\text{SOT}$. We further observe a striking reduction of the total switching time from 15\,ns to less than 1\,ns at $V_\text{SOT} \geq 1.7\,V_\text{c}$, which can be interpreted as switching at sub-ns timescale in the great majority of the 1000 trials, with the confidence given by the signal-to-noise ratio.

\begin{figure}[b]
\includegraphics{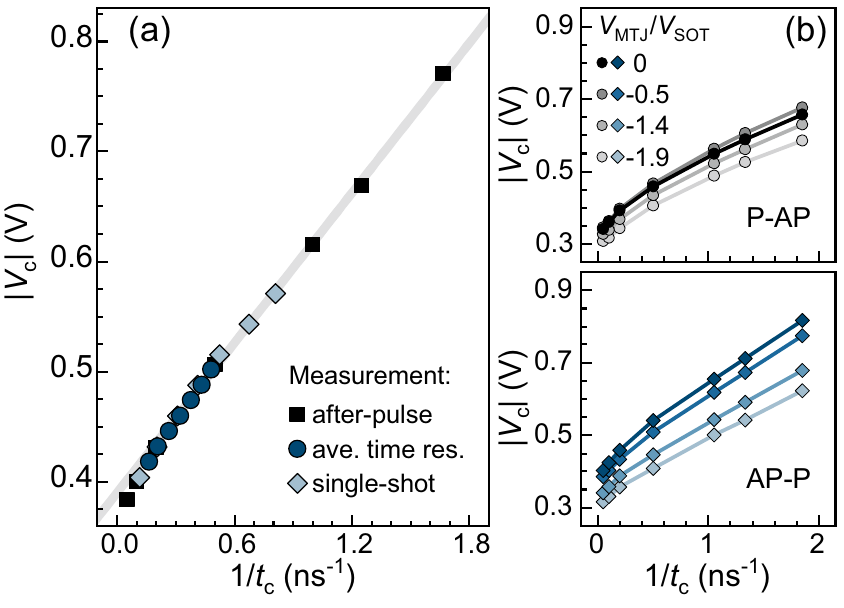}
\caption{\label{fig4:Vc} (a) $V_\text{c}$ obtained from after-pulse probability (squares), real-time single-shot measurements (diamonds), and averaged time-resolved measurements (circles). The gray line represents a fit to the data by a $1/t_\text{c}$ function for $t_\text{c}<4$\,ns. (b) $V_\text{c}$ for SOT-dominated switching at different values of $V_\text{MTJ}$ obtained from after-pulse probability measurements for P-AP (top panel) and AP-P (bottom panel) switching.}
\end{figure}

Our measurements also allow us to compare the critical switching voltage obtained by after-pulse, single-shot, and averaged time-resolved measurements. This comparison is relevant to assess the reliability of different methods employed to measure $V_\text{c}$, particularly for the more common after-pulse resistance measurements, in which $t_\text{0}$ or $\Delta t$ cannot be accessed. We define the critical switching time ($t_\text{c}$) as the pulse width corresponding to 50\% switching probability in after-pulse measurements, and as $t_\text{0}+\Delta t/2$ and the time required to reach $V_\text{sw}$~=~0.5, in single-shot and averaged time-resolved measurements, respectively. Likewise, we define $V_\text{c}$ as the corresponding SOT pulse amplitude. Figure\,\ref{fig4:Vc}(a) shows that, irrespective of the measurement method, all values of $V_\text{c}$ fall on the same curve and scale inversely with the critical time $t_\text{c}$ for pulses shorter than 4\,ns. Such a scaling is expected for the intrinsic regime, in which conservation of angular momentum gives $1/t_\text{c}\propto (V_\text{c}-V_\text{c0})$ \cite{Bedau2010,Liu2014a,Garello2014}. Here $V_\text{c0}$ is the intrinsic critical voltage that reflects the minimum amount of angular momentum required to achieve switching in the absence of thermal effects. A linear fit of the data in Fig.\,\ref{fig4:Vc}(a) gives $V_\text{c0}$\,=\,390\,mV, which corresponds to an intrinsic critical current of $1.4\times 10^{12}$\,A\,m$^{-2}$. For pulses longer than 4\,ns, deviations from the linear behavior are attributed to the onset of thermally-activated switching \cite{Bedau2010,Liu2014a,Garello2014}.

Figure\,\ref{fig4:Vc}(b) shows that in both switching configurations, $V_\text{c}$ reduces considerably upon increasing $V_\text{MTJ}$. The reduction of $V_\text{c}$ is largest for the shorter pulses and for AP-P switching (bottom panel) compared to P-AP switching (top panel). These observations can be explained by the combined impact of VCMA and heat generated by the bias. $V_\text{MTJ}>0$ facilitates switching by weakening the anisotropy of the free layer, whereas the bias current simultaneously increases the temperature in the device, thus lowering the switching energy barrier regardless of its sign. Therefore, when $V_\text{SOT}$ is negative (AP-P switching), both VCMA and temperature lead to a reduction of $V_\text{c}$; however, when $V_\text{SOT}$ is positive (P-AP switching), the VCMA opposes the temperature-induced decrease of the switching barrier, resulting in a smaller reduction of $V_\text{c}$. Consequently, $V_\text{c}$ in both switching configurations equalizes for $V_\text{MTJ} \approx -1.65 V_\text{SOT}$. This result shows that $V_\text{MTJ}$ can be efficiently used to realize symmetric switching conditions [see also Fig.\,\ref{fig2:TRsingles}(e)].

\begin{figure}
\includegraphics{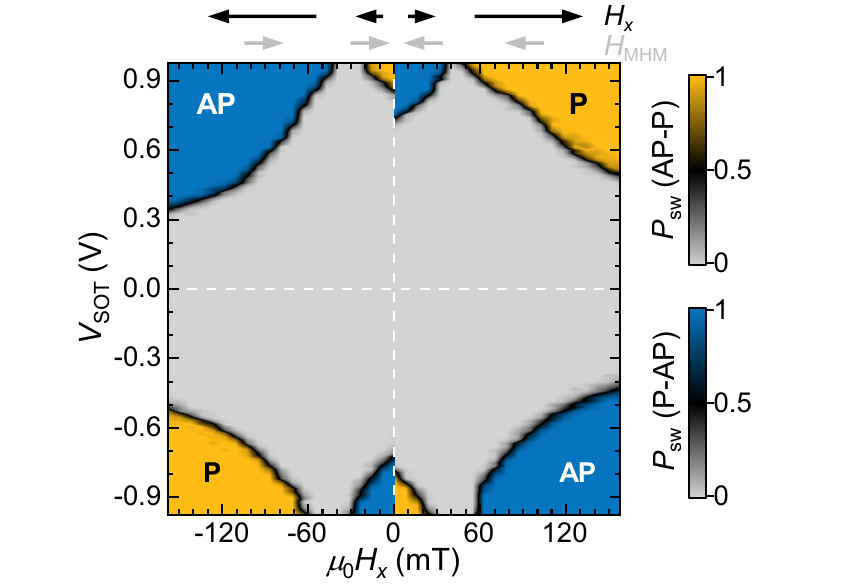}
\caption{\label{fig5:Psw} SOT switching probability as a function of $V_\text{MTJ}$ and $\mu_0 H_x$ for 0.5\,ns-long pulses at $V_\text{MTJ}=0$. The color of each point represents the after-pulse switching probability $P_\text{sw}$ (out of 50 trials). The magnetization of the MHM was initially set to be negative (positive) for $H_x <0$ ($>0$), resulting in $H_\text{MHM}>0$ ($<0$) as indicated by the arrows above the diagram. The AP/P labels denote the final state.}
\end{figure}

Last, we address the functionality of our devices under an external field ($H_x$) by measuring the after-pulse switching probability ($P_\text{sw}$) for different pulse amplitude and field conditions. We start by considering SOT switching at $V_\text{MTJ}$\,=\,0. Each point in the switching phase diagram in Fig.\,\ref{fig5:Psw} illustrates the statistical result of 50 trials for a fixed pulse width (0.5\,ns). The black boundary defines $P_\text{sw}$\,=\,0.5, which divides the diagrams into under- (gray) and over-critical (blue and yellow) regions. Since $H_x$ polarizes the MHM, which has a coercivity of about 20\,mT, $H_\text{MHM}$ is always antiparallel to $H_x$ [Fig.\,\ref{fig1:SampleSetup}(a)]. As a consequence, the switching polarity depends on the sign of $|H_\text{MHM}-H_x|$, i.e., of the total in-plane field acting on the free layer. The diagram also allows for evaluating the strength of $H_\text{MHM}$ from the difference between two $H_x$ values resulting in the same $V_\text{c}$. In this manner, we estimate that $\mu_0 H_\text{MHM}\approx 40$\,mT. The diagram shows that bipolar switching is possible in a wide range of external fields with the exception of narrow intervals, in which $|H_\text{MHM}-H_x|$ approaches zero. A typical transition of $P_\text{sw}$ from 0.01 to 0.99 occurs upon increasing $V_\text{SOT}$ by less than 80\,mV, in contrast with STT switching, for which a 150\,mV increase of $V_\text{MTJ}$ is required using the same device and 15\,ns-long pulses.

Before concluding, we discuss a few possibilities to improve the switching speed and the design of field-free SOT devices. As shown in Ref.~\onlinecite{Grimaldi2020}, increasing the magnitude of the in-plane field significantly reduces the switching time for a given $V_\text{SOT}$. In general, $H_\text{MHM}$ can be increased by i) optimizing the aspect ratio and thickness of the magnetic layer, ii) replacing the Co layer by a material with higher saturation magnetization, such as CoFe, and iii) bringing the MHM closer to the MTJ. The optimal strength of $H_\text{MHM}$ will ultimately depend on the critical current, switching rate, and thermal stability of the free layer set by the target application. Static measurements show that the TMR is not affected by the MHM and that the overall device properties are more influenced by the design of the MTJ pillar rather than by the hard mask itself \cite{Garello2019}. Better compensation of the out-of-plane stray field produced by the SAF would lead to a more symmetric switching behavior between the P and AP configurations. This can be achieved, e.g., by changing the thickness of the reference layer or the thickness and number of repetitions of the SAF multilayer. Alternatively, as shown here, $V_\text{MTJ}$ can be used to balance the SAF field. The MHM approach is also compatible with dense designs, as shown in Ref.~\onlinecite{Garello2019}. Micromagnetic simulations further show that surrounding magnets have a stabilizing effect on the magnetization of the hard mask. MHMs down-scaled to a volume of $50\times 100\times 25$\,nm$^3$ and a pitch of $100\times150$\,nm$^2$ have more uniform magnetization patterns than MHMs with a volume of $110\times390\times50$\,nm$^3$ and a pitch of $260\times 540$\,nm$^2$. Future studies might establish if the MHM approach based on a single mask per MTJ is compatible with sharing the same SOT write line between multiple MTJs, as proposed in Ref.~\onlinecite{Kato2018a}.

In summary, we have demonstrated field-free switching of perpendicularly magnetized MTJs by SOT in real time. Single-shot time-resolved measurements show that the stochastic incubation delay near the critical voltage threshold ($V_\text{SOT} \approx V_\text{c}$) is several ns long for both the AP-P and P-AP switching configurations, whereas the actual transition time is about 1\,ns. Upon increasing $V_\text{SOT}$ or $V_\text{MTJ}$, the switching distributions narrow down leading to reduced latency and quasi-deterministic switching. Averaged time-resolved measurements show that the total switching time can be reduced to 0.7\,ns by increasing $V_\text{SOT}$ up to $1.9 V_\text{c}$ with negligible assistance of either STT or VCMA. At timescales shorter than 4\,ns, the critical switching voltage is found to scale linearly with inverse of the switching time, as expected in the intrinsic regime. Real-time measurements and after-pulse switching statistics as a function of pulse length are found to provide a consistent estimate of the critical switching time. Measurements of the switching probability as a function of $V_\text{MTJ}$ and external field indicate that further improvements of the switching dynamics and reduction of $V_\text{c}$ can be obtained by VCMA, increase of the dipolar field of the MHM, and compensating the dipolar field due to the SAF.\newline


This research was supported by the Swiss National Science Foundation (Grant no. 200020-172775), the Swiss Government Excellence Scholarship (ESKAS-Nr. 2018.0056), the ETH Zurich (Career Seed Grant SEED-14 16-2) and imec’s Industrial Affiliation Program on MRAM devices.\newline

This article may be downloaded for personal use only. Any other use requires prior permission of the author and AIP Publishing. This article appeared in Appl. Phys. Lett. \textbf{116}, 232406 (2020) and may be found at \href{https://doi.org/10.1063/5.0011433}{https://doi.org/10.1063/5.0011433}.


\bibliography{Literature}

\end{document}